# Block Chain based Intelligent Industrial Network (DSDIN)

Barco You, Matthias Hub, Mengzhe You, Bo Xu, Mingzhi Yu and Ivan Uemlianin

*Dasudian Technologies Ltd.*

*Abstract*—**The manufacturing industry featured centralization in the past due to technical limitations, and factories (especially large manufacturers) gathered almost all of the resources for manufacturing, including: technologies, raw materials, equipment, workers, market information, etc. However, such centralized production is costly, inefficient and inflexible, and difficult to respond to rapidly changing, diverse and personalized user needs. This paper introduces an Intelligent Industrial Network (DSDIN), which provides a fully distributed manufacturing network where everyone can participate in manufacturing due to decentralization and no intermediate links, allowing them to quickly get the products or services they want and also to be authorized, recognized and get returns in a low-cost way due to their efforts (such as providing creative ideas, designs or equipment, raw materials or physical strength). DSDIN is a blockchain based IoT and AI technology platform, and also an IoT based intelligent service standard. Due to the intelligent network formed by DSDIN, the manufacturing center is no longer a factory, and actually there are no manufacturing centers. DSDIN provides a multi-participation peer-to-peer network for people and things (including raw materials, equipment, finished / semi-finished products, etc.). The information transmitted through the network is called Intelligent Service Algorithm (ISA). The user can send a process model, formula or control parameter to a device via an ISA, and every transaction in DSDIN is an intelligent service defined by ISA.**

## I. Introduction

AT present, a new round of scientific and technological revolutions and industrial reforms is coming, and the global industrial Internet is under rapid development. As a core carrier for constructing the industrial Internet ecosystem, the industrial Internet platform is expanding from the commercial field to the manufacturing industry, becoming a driving force to combine manufacturing with the Internet and playing a key role in strategic layout for countries, industries and leading companies in the world. The industrial Internet platform is designed for the manufacturing industry in need of digitalization, networking and intelligentization. It builds a carrier based on mass data collection, aggregation, analysis and service systems and supporting the ubiquitous connection, flexible supply and efficient configuration of manufacturing resources. Its core elements include data collection system, industrial PaaS and application service system. Data collection system: the data about equipment, systems, products and the like are collected through such technologies as smart sensors, industrial control systems, IoT technologies and intelligent gateways. Industrial PaaS: the platform combines cloud computing, big data and experience accumulated in practical industrial production to form basic industrial data analysis capability; technology, knowledge, experience and other resources are materialized into transplantable and reusable software tools and development tools like special software library, application model library and expert knowledge base to build an open and shared development environment. Application service system: it is designed for such industrial needs as asset management optimization, technological process optimization, manufacturing collaboration and resource sharing, and provides users with intelligent applications and solutions.

The industrial Internet platform plays an important role in creating a new type of industry and promoting the combination of "Internet + advanced manufacturing" for the following reasons:

1) It can give play to the agglomeration effect of the Internet platform. The industrial Internet platform is a carrier of the data about hundreds of millions of equipment, systems, process parameters, software tools, enterprise business needs and manufacturing capabilities for industrial resources sharing. It plays a key part in networking, synergy and optimization, and results in a series of new Internet business models, including the manufacturing-based crowdsourcing and creative space, collaborative manufacturing and intelligent services.

2) It plays a key role in the industrial operating system. On one hand, the industrial Internet platform connects with numerous devices and works as a carrier of industrial experience and knowledge models; on the other hand, it links to industrial optimization applications and plays a key role in connecting all industrial elements and allocating industrial resources and contributes to the development of advanced intelligent manufacturing systems.
3) It can give full play to the cloud computing as well as edge computing [1] resources. Due to advanced architecture and high-performance infrastructure for cloud computing and edge computing, the industrial Internet platform can make possible the integration, storage and computation of massive heterogeneous data to provide solutions to the contradiction between explosive growth of industrial data to be processed and the existing system's disadvantage in computing power, and accelerate the process of networking and intelligentization driven by data.

Many manufacturing companies have started to provide services for customers and tried to increase the proportion of services in the total income, but most of them are still providing traditional services based on their products, such as after-sales service, product leasing, purchasing products for customers, and providing financing services etc. These traditional services can bring customers limited value and often fail to keep up with customers' needs. Therefore, it is difficult for enterprises to realize service transformation by only providing traditional services. However, the emergence of the Internet of Things (IoT) brings a new opportunity for enterprises which struggle for service transformation. As mentioned above, the changes brought by Industry 4.0 are all data-driven [2]. The IoT obtains data from the physical world through various sensors, and then analyzes and uses these data to help enterprises optimize their production processes and improve their operating efficiency [3]; more importantly, with help of the IoT, these enterprises can be well informed of customers' needs and create new service models to promote their business growth. This is the greatest value of the IoT for enterprises [4].

Enterprises can provide their customers with real-time and personalized intelligent services by using the data generated by the IoT [5]. In essence, these services are different from traditional after-sales services in the fact that the data collected via the IoT can be used to predict customers' needs in a real-time and sustainable way and the system will automatically optimize and adjust the services according to the analysis result and can even automatically adapt to the environment, independently make decisions and bring highly personalized experience for customers [6][7]. For example, an equipment manufacturer can predict via a sensor installed in the equipment that some component of customer's equipment needs to be replaced, and transport the spare part to a warehouse near the customer in advance, which can greatly shorten the time for the customer to wait for the replacement and reduce the loss arising from shutdown; and on the other hand, the customer has no need to hoard a large number of spare parts, which occupies funds and warehouses; this method can also reduce the possibility of customers to purchase spare parts from other brands and increase the manufacturer's income. Enterprises can also create new service models through the IoT, such as sharing their own manufacturing capabilities and providing production services for other enterprises. In this way, everyone can become a main role in manufacturing, either demander or participant; you can provide C2B customized services according to the customers' needs; you can also provide them with financing and insurance services based the data in the IoT. Those leading enterprises in their own industry can also build an IoT based platform and become the center of the industry ecosystem [8][9]. These new opportunities to create values brought by the IoT for the manufacturing industry need to be combined with different industries and other emerging technologies, including data analysis, artificial intelligence and blockchain to form real intelligent services and create values [10] [11].

The key to the Industry 4.0 is the informatization of raw materials (substances), i.e. combining the so-called physical world with the cyberspace to bring about a real cyber-physical system (CPS) [12]. Specifically, the raw materials purchased by a factory and the finished/semi-finished products are "labeled", for example: This is ×× product made of ×× materials based on ×× technologies for Customer A. In other words, the "raw materials" containing information are used by the smart factory to make possible the shift from "material flow" to "information flow". Materials and their production and processing have seen value delivery via the IoT [8] [13]. Finally, the manufacturing industry is bound to become part of the information industry, so the Industry 4.0 is the last industry revolution [14].

To overcome the above challenges and grasp the opportunities for development not only depends on corporate capabilities and internal resources. First of all, the data collected via the IoT is meaningless if they do not combine with different industries and emerging technologies such as data analysis, artificial intelligence, blockchain, cloud computing and fog computing. Only such combination can result in innovative services. Moreover, due to the fact that intelligent services have numerous application scenarios and feature the real-time and dynamic development, an enterprise cannot satisfy its customers' needs if only depending on its own resources. Due to the uncertainty of demands for intelligent services and the uncertainty of return on investment, the enterprise should cooperate with other companies to share these risks. Therefore, to develop intelligent services, the enterprise should work closely with relevant external partners to form a value creation network. The IoT based intelligent services will go across the boundary between corporate value chain and traditional industry to create a new ecosystem, changing the existing competitive landscape and breaking the inherent rules for victory. We call this value creation network as "Block Chain based Intelligent Industrial Network (DSDIN)"

We have seen the Bitcoin based blockchain technology combined with a new way of value exchanges on the Internet [15], leading to some promising progress: Ethereum demonstrates that Turing's complete smart contract can be realized by a blockchain architecture [16]; Truthcoin creates tools to make oracle possible in the blockchain [17]; Casey Detrio shows how to build markets on the blockchain [18]; Namecoin creates a distributed version of the Domain Name Service (DNS) [19]; Factom shows how to prove the existence of any digital asset by using hash values stored in the blockchain [20].

Based on the distributed computing technology – Erlang [21][22], DSDIN fully combines the IoT with blockchain technologies to build a network of mutual trust for interaction of industrial data and intelligent services.

## II. DESIGN PRINCIPAL

This Paper starts with the key problems facing the current development of our industrial Internet platform and make efforts in both sides of supply and demand to build the DSDIN with reference to the the "Internet + Advanced Manufacturing Industry" guideline to develop the Industrial Internet. We take into full consideration the future development trend, focus on the key points of such combination, highlight the ecosystem building of "IoT + AI + blockchain", put emphasis on enhancing the foundations for development and improving the platform's operating capacity and provide support for industrial enterprises in their transformation for digitization, networking and intelligentization to realize interconnection of internal departments and upstream and downstream industries, cross-field production equipment and information system, thus getting rid of "information island" and contributing to integration and sharing of resources and data. We should allow such service components as process models, knowledge components, algorithm tools and development tools to be shared by third-party developers who will be guided to develop new industrial intelligent applications and services based on the DSDIN platform, forming an IoT based innovative ecosystem for industrial intelligent service developers. With the help of the platform which is a carrier of massive data about equipment and products, we should make great efforts to build a favorable community for intelligent service developers and use the platform-wide available Token (DSD) to encourage industrial enterprises and third-party developers to use the common models, knowledge components, algorithm tools, development tools, and operating environments on the platform and to develop intelligent service applications and algorithm models and apply algorithm models to realize intelligent services. Additionally, these efforts can improve the efficiency of producing, spreading and reusing the industrial knowledge to form a virtuous cycle of mutual promotion and bidirectional iteration between the improved platform and massive applications and to accelerate the formation of a standard industrial intelligent network.

DSDIN is a blockchain based IoT and AI technology platform, and also an IoT based intelligent service standard. Due to the intelligent network formed by DSDIN, the manufacturing center is no longer a factory, and actually there are no manufacturing centers. DSDIN provides a multi-participation peer-to-peer network for people and things (including raw materials, equipment, finished / semi-finished products, etc.). The information transmitted through the network is called Intelligent Service Algorithm (ISA). The user can send a process model, formula or control parameter to a device via an ISA, for example: When a water cup designer sends his design model to a 3D printer through an ISA, the printer obtains corresponding raw materials according to the ISA, prints the cups at the agreed quantity according to the preset parameters, and sends them to the designated address by express delivery. Every transaction in DSDIN is an intelligent service defined by ISA.

### A. DSDIN Blockchain

The DSDIN network consists of a public blockchain and a myriad of independent state machines (also known as wormholes). The wormholes surround the blockchain and result in a large number of state channels, as shown in Fig. 1. In the blockchain design for DSDIN, we believe that it is not necessary to maintain the state in the blockchain. We only need to store the state information on the wormhole and use the blockchain to deal with the economic consequences of any information exchange, including money transfer, mining, etc., as well as the fallback in case of dispute. So, in the DSDIN network, we propose another blockchain architecture that places Turing's complete smart contract on the wormhole, rather than the blockchain. This increases the extensibility of the system and the throughput of the transaction, because a wormhole is an independent state machine (a single process or process tree based on DSDIN VM), making all transactions become independent and enabling parallel processing. In addition, this also means that the contract is not written to the shared state, greatly simplifying the testing and verification. At the same time, this separates the economic logic from the data storage, allowing us to supplement the blockchain with a good distributed storage scheme, improving the efficiency, privacy and security of data storage in the physical world.

DSDIN is similar to Ethereum, which is based on scripts, competing coins and on-chain meta-protocols and state channels for integration and enhancement, enabling developers to create arbitrary consensus-based, extensible, standardized and fully featured collaborations that are easy to develop and able to connect with the real-world data applications. By establishing the ultimate and abstract base layers and the wormholes with the Turing's complete programming language, DSDIN enables anyone to create contract and decentralize applications and set up their freely defined ownership rules, trading methods and state transition functions. Smart contracts can also be created and executed on wormholes, and because of Turing completeness, value-awareness, blockchain-awareness, and multi-state collaboration, these contracts are more powerful than those provided by Ethereum. For contract developers, the DSDIN blockchain is "stateless" but also tracks predefined state components.

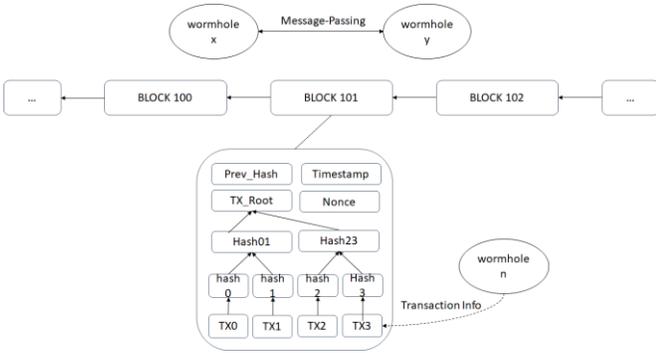

Fig. 1. The blockchain and wormhole in DSDIN. Note that wormholes communicate with each other by Message-Passing and only the information regarding transactions happening in wormholes is stored onto blocks.

*B. Blocks*

In the DSDIN's blockchain, each block contains the followings:
- The hash value of the previous block
- The Merkle tree for transactions
- The Merkle tree for accounts
- The Merkle tree for the naming system
- The Merkle tree for wormholes
- The Merkle tree for oracles that have not yet provided an answer to the question
- The Merkle tree for answers provided by oracles
- The Merkle tree of Merkle proofs
- Entropy of random number generator

Among them, the hash value of the previous block is used to keep the chain order. The transaction tree contains all transactions in this block. All trees are consistent. If a tree changes from one block to another, such change must be caused by an additional transaction made on the transaction tree of the new block. The Merkle proof tree must contain an updated Merkle proof.

*C. Consensus Mechanism*

The DSDIN network uses the consensus mechanism of Proof-of-Work (PoW) and Proof-of-Stake (PoS), and the order of blocks depends on PoW. The Tromp's Cuckoo Cycle and the derives will be used as the PoW mechanism for DSDIN. It is a memory bound and indirect PoW mechanism based on graph theories, featuring less power consumption. As described by Tromp: Cuckoo Cycle is a PoW mechanism that is memory bound and can be certified instantly; its uniqueness lies in the fact that it depends on memory latency, not computation to achieve PoW [23]. Therefore, the Cuckoo Cycle based mining can be an ASIC mining, and DRAM is only used as a support for byte read-out and write-in. Even fully charged mobile phones can be used for mining, having no impact on the efficiency of mining.

The consensus mechanism is a non-standard configuration in DSDIN, used for agreeing on addition of a new block and also for answering the questions proposed by oracles and defining the system parameters, which helps the system adapt to the environment changes and the development of new technologies. In addition, the consensus mechanism can also revise and update itself. If the PoW mechanism is too simple, it is easy for miners to win the oracle, so DSDIN will use an innovative algorithm combining PoW and PoS to give full play to their own advantages. At the same time, PoW will be used to release new DSD Tokens.

Some Ethereum based applications such as Augur, try to use a blockchain for distributed storage of the real-world data. They use a smart contract to build a consensus mechanism instead of directly using the consensus mechanism in the blockchain, which results in low efficiency but no enhancement in data security. In order to improve efficiency, we need to build a consensus mechanism that provides not only the information about network state, but also about the state of the outside world. This consensus mechanism helps us to get a real oracle and can answer the questions that Turing might fail to answer, such as: The100th product I produce is the most satisfying one for customers?

*D. Wormhole*

The wormholes in DSDIN are the carriers to enable state channels which are used to ensure that only relevant entity in a transaction have access to the information and data about the transaction. Essentially, the participants of the transaction instantiate some states onto the blockchain, such as Ethereum contract and bitcoin multisig [15] [16], and then they only need to send the updated signature to each other. Importantly, each participant can use such information to update the state of a blockchain, but they will not do so in most cases. This allows a transaction to be implemented at the same speed as information spreading, without waiting for the transaction to be verified and finally confirmed by the consensus mechanism of the blockchain.

In DSDIN, only the state of DSD Token transfer can be updated via a blockchain, and only DSD Token stored in a wormhole by any transaction party can be transferred. This requires each wormhole to be independent of each other, so that the transactions in each wormhole can be processed in parallel, greatly improving the transaction throughput. Blockchains are only used for final clearing and conflict resolution. Since the behavior of a blockchain is predictable, the dispute over the outcome of running wormhole will bring no benefit, so malicious participants are encouraged to act correctly and the blockchain is used for final settlement. All of these measures increase the speed and size of transactions and also result in better privacy.

As shown in Fig. 1, each wormhole is a stand-alone virtual machine (DSDIN VM) that implements smart contracts, ISA and actual transactions through state channels, with the transaction information finally recorded on the blockchain. Each wormhole is similar to a process in a computer, protecting its own state data, including: Data generated during running, data in the database and data acquired through interfaces (such as HTTP, TCP, Email, Serial Port etc.). The wormholes exchange information with each other via Message-Passing but do not share data so as to keep their independence, similar to the standard Actor-Model program [24][25][26]. This allows transactions in the entire system and ISA to realize high-throughput running and also guarantees the security of private wormhole data.

Each wormhole corresponds to a process tree. As shown in Fig. 2, a wormhole consists of Supervisors and Workers. Each Work performs independent distributed computing, such as enabling a state channel to manage the real-world data (such as read-write of data via a database or communication with a CNC machine etc.) and exchange information with other wormholes. Supervisor generates Workers, and when Workers fail, Supervisor can restart Workers based on certain strategies [27].

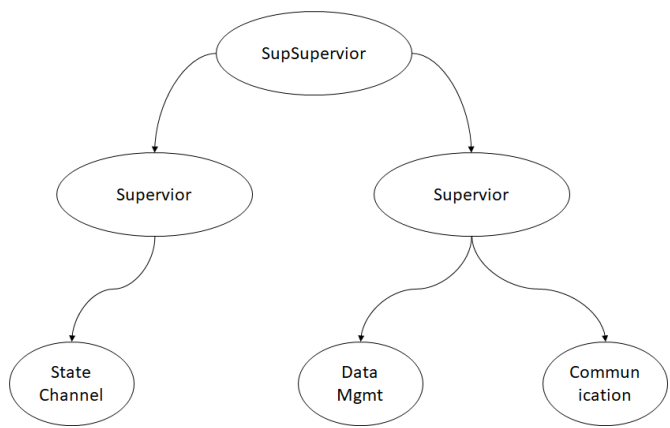

Fig. 2. The process tree corresponding to a wormhole in DSDIN.

### E. Smart Contract

Although only the state of the DSD Token transfer can be updated on a blockchain, DSDIN still runs a Turing-complete Smart Contract Virtual Machine (DSDIN VM), and the contract defines the rules for fund allocation in DSDIN. Unlike Ethereum's physical contract, on one hand, only the participants are informed of the existence of the contract; on the other hand, only those participants on the wormhole that has public status channels can create a legal contract. If the participants agree on the contract, the contract will be signed and saved forever. Only in the event of a dispute outcome will the contract be submitted to the blockchain, in which case only part of codes containing information about the ever-submitted transaction will be submitted and the blockchain will distribute Token according to the contract and close the state channel on the wormhole.

More importantly, in DSDIN, a contract does not contain its own state, so any state is maintained by the participants of a transaction, and these states will be submitted as input parameters when the contract is executed. Each contract is actually a pure function, taking full advantage of the ideas for functional programming, which obtains the input parameters and outputs new channel states.

In DSDIN network, the quantitative execution method of the contract is similar to the "gas" of Ethereum. However, we measure with two different resources: one is for time and the other is for space. They both require participants of the contract executed to pay with DSD Token. For example, if Alice and Bob want to transact with the wormhole on the DSDIN network, they need the following steps:

1. Alice and Bob sign a contract respectively than specifies the quantity of money deposited in the wormhole and publish to the blockchain.
2. Once the blockchain opens this channel, they can both create new channel status and send and sign each other.

The channel status can be a new fund distribution, or a contract related to the new distribution. Each channel status has an incremental nonce and is signed by both parties. If a dispute arises, the latest valid status will be submitted to the blockchain finally for compulsory execution.

The channel can be closed in two ways: a) if Alice and Bob decide to complete the transaction and agree to their final account balance status, they shall sign the transaction and submit it to the blockchain. So, the channel is closed and the money on the channel is correspondingly redistributed; b) If Alice refuses to sign or wants to close the transaction for any reason, Bob can submit the last status they agree on and sign to the blockchain and ask to close the channel in this status. This will trigger a countdown. If Alice thinks Bob is dishonest, she will have the opportunity to release a larger nonce that both of them have signed before the end of the countdown. If Alice does this, the channel will be immediately closed, otherwise it will not be closed until the countdown ends.

### F. Transaction

Transactions on DSDIN are implemented through smart contracts and the process is as follows:

1. Check whether the format of the transaction is correct (i.e., with the correct value), whether the signature is valid, and whether the random number matches the random number of the sender's account. If it is not, an error will be returned.
2. The sender deposits the Deposit.
3. Calculate the transaction fee: fee=Gas * GasPrice; and determine the sender's address from the signature. Subtract the transaction fee from the sender's deposit and increase the sender's random number. If the deposit balance is insufficient, an error will be returned.
4. Transfer the value from the sender's account to the recipient's account. Create this account if the receiving account does not yet exist. If the receiving account is a contract, run the contract code until the code running ends or the gas is used up.
5. If the value transfer fails due to insufficient deposit saved by the sender or running out of the gas during code execution, the original status will be restored. However, the transaction fee is also required, and the transaction fee is added to the miner's account.
6. Otherwise, all remaining gas will be returned to the sender, and the consumed gas will be sent to the

miner as the transaction fee.

If there is no contract to accept the transaction, then all transaction costs are equal to GasPrice multiplied by the byte length of the transaction, and the transaction data is independent of the transaction cost. In addition, it should be noted that the contract-initiated message can allocate a gas limit to the calculations they generate. If the sub-calculated gas is used up, it will only return to the status when the message was sent. Therefore, just like a transaction, a contract can also protect their computing resources by setting strict limits on the sub-calculations it generates.

The code for creating a contract transaction is shown in Fig. 3, 4 and 5.

```
new(#{owner      := OwnerPubKey,
      nonce      := Nonce,
      code       := Code,
      vm_version := VmVersion,
      deposit    := Deposit,
      amount     := Amount,
      gas        := Gas,
      gas_price  := GasPrice,
      call_data  := CallData,
      fee        := Fee}) ->
    {ok, #contract_create_tx{owner     = OwnerPubKey,
                             nonce     = Nonce,
                             code      = Code,
                             vm_version = VmVersion,
                             deposit   = Deposit,
                             amount    = Amount,
                             gas       = Gas,
                             gas_price = GasPrice,
                             call_data = CallData,
                             fee       = Fee}}.
```

Fig. 3. Create a new transaction in DSDIN.

```
check(#contract_create_tx{owner = OwnerPubKey, nonce = Nonce,
                          fee = Fee}, Trees, Height) ->
    Checks =
        [fun() -> dsdin_utils:check_account(OwnerPubKey, Trees, Height, Nonce, Fee) end],
    case isa_validation:run(Checks) of
        ok            -> {ok, Trees};
        {error, Reason} -> {error, Reason}
    end.
```

Fig. 4. Check transaction legality.

### G. Oracle

The oracle represents the strict conditions defined in the contract to interact with the environment, such as the price of different commodities or the occurrence of a particular event (such as the malfunction of parts of a machine). If the smart contract does not contain these features that interact with the outside world, the contract will be closed and useless. So, in the DSDIN network, we try to introduce external real-world data into the blockchain in a distributed way. However, for most blockchains, to determine whether the facts provided are correct or not, a new consensus mechanism (such as Ethereum's smart contract) needs to be implemented based on its own consensus mechanism. Running two sets of consensus mechanisms inter-lapped with each other is extremely costly than running them independently, and there is no slight increase in the security. Therefore, we combine the two-layer consensus mechanism into one, essentially reusing the consensus mechanism for the

```
process(#contract_create_tx{owner = OwnerPubKey,
                            nonce = Nonce,
                            fee   = Fee} = CreateTx, Trees0, Height) ->
    AccountsTree0  = isa_trees:accounts(Trees0),
    ContractsTree0 = isa_trees:contracts(Trees0),

    %% Charge the fee to the contract owner (caller)
    Owner0      = isa_accounts_trees:get(OwnerPubKey, AccountsTree0),
    {ok, Owner1} = isa_accounts:spend(Owner0, Fee, Nonce, Height),
    AccountsTree1 = isa_accounts_trees:enter(Owner1, AccountsTree0),

    %% Create the contract and insert it into the contract state tree
    %% The public key for the contract is generated from the owners pubkey
    %% and the nonce, so that no one has the private key. Though, even if
    %% someone did have the private key, we should not accept spend
    %% transactions on a contract account.
    ContractPubKey = create_contract_pubkey(OwnerPubKey, Nonce),
    Contract       = dsdin_contracts:new(ContractPubKey, CreateTx, Height),
    ContractsTree1 = dsdin_state_tree:insert_contract(Contract, ContractsTree0),

    Trees1 = isa_trees:set_accounts(Trees0, AccountsTree1),
    Trees2 = isa_trees:set_contracts(Trees1, ContractsTree1),

    {ok, Trees2}.
```

Fig. 5. Processing the execution of a transaction.

status of the system. The approval of the system status is also the approval of the external world.

Specific mechanism as follows: The owner of any DSD Token can run an oracle, focus on answering yes/no questions, and specify the time window in which the question is answered. This time window can start now or at some point in the future. The number of DSD Token deposited for the user to turn on the oracle is directly proportional to the length of the time window. If the user gets a correct answer within the time window, the remaining DSD Token will be returned to the user account, otherwise it will be burned. The blockchain will generate a unique identifier for each oracle to obtain the answer to the question. Once it is time to answer the question, the user starting the oracle can answer the question for free. If the operator of the oracle gives an answer or the time window of the question expires, other users can file a counterclaim by depositing the same number of DSD. If no counterclaim is filed within the time window, the answer provided by the oracle operator is accepted as the correct answer and the deposit will be returned. Conversely, if someone raises a counterclaim, then the block consensus mechanism will be used to answer the oracle.

### H. Token of the DSDIN Network (DSD)

DSD is the primary encryption fuel in the DSDIN network to offset transaction costs, and users consume services and resources in the DSDIN network by consuming DSD. DSD takes its name from Manufacturing Intelligence, which means that DSD is consumed in exchange for manufacturing intelligence. The economic application realized through the platform, including transfer, sub-accounting, payment, data exchange etc., requires consumption of DSD. The founding block for the first release of DSD is defined by a genesis smart contract, and more DSDs will be produced through mining.

### I. DSDIN Account

In the Ethereum system, the status consists of an object called the "account" (each account is a 20-byte address) and a status transition that transfers value and information between two accounts. DSDIN's account consists of four parts:

- Account address

- DSD balance
- A counter that will be incremented per transaction
- The freshness of the last update

There are two types of accounts in the DSDIN network: external accounts (controlled by the private key) and contract accounts (controlled by the contract code). All external accounts have no code, and people can send messages from an external account by creating and signing a transaction. Whenever a contract account receives a message, the code inside the contract will be activated, allowing it to read and write to the internal storage, and to send other messages or create contracts.

Each account needs to consume a small amount of account maintenance fees since it was created. When the account is deleted, it can receive some rewards to encourage more space released on the platform.

*J. Naming System*

For many blockchains, the user's address is a complex Hash character string that is unreadable. While in the DSDIN network, we provide a distributed and secure naming system to support human-readable and easy-to-remember names. The status of the blockchain contains a unique correspondence that maps readable names to the fixed-length string spaces. These names can be used to point to an account address, or a Hash value, such as a node in the Merkle tree.

*K. Decentralized Storage*

In the past, some cloud-based online file storage services, such as Dropbox, allow users to upload their hard drive backups, provide backup storage services and allow user access and charge users on a monthly basis. However, at this point, this file storage market is sometimes relatively inefficient. DSDIN's smart contracts allow decentralized storage ecosystem development, so users can reduce the cost of file storage by leasing their own hard drives or unused network space for a small amount of revenue. The working principle of the contract is as follows: First, someone divides the data that needs to be uploaded into blocks, encrypts each block of data to protect privacy, and then builds a Merkel tree. Then this person shall create a contract with the following rules. For every N blocks, the contract will extract a random index from the Merkel tree (using the hash of the previous block that can be accessed by the contract code to provide randomness), and then give the first entity a certain amount of DSD to support a proof of ownership of a block at a particular index in the tree with a similar simplified verification payment (SPV) [15]. When a user wants to re-download his file, he can use the micro-payment channel protocol (for example, 0.1DSD per 64K bytes) to recover the file; for the expense, the most efficient way is that the payer does not release the transaction until the end, and shall use a slightly more cost-effective transaction with the same random number to replace the original transaction after every 64K bytes.

An important feature of this protocol is that although it seems that a person trusts many random nodes that will not lose files, he can divide the file into many small pieces by secret sharing, and then monitor the contract to know that each small piece is still kept by certain node. If a contract is still paying, then evidence is provided that someone is still saving the file.

*L. Edge Intelligence*

Wormholes in the DSDIN network can also be used to create a verifiable computing environment that allows users to invite others to perform computations, run an ISA (such as real-time control of a CNC machine), and then selectively request to provide the evidence that the computation is correctly completed at a randomly selected checkpoint. This allows any users to participate in the implementation of the ISA with their desktops, laptops, gateways or industrial personal computers in the production line. On-site inspection and security deposits can be used to ensure that the system is trustworthy (i.e. no nodes can benefit from deception), while also ensuring the interaction between both parties of the transaction on the basis of no trust to achieve physical manipulation (such as 3D printing) and ISA algorithm. When the computation result is proved at a certain randomly selected checkpoint, the smart contract will be automatically performed, realizing fund settlement, data key expiration, etc.

*M. Extensibility*

The architecture of the DSDIN network is highly extensible. If each user (node) only saves some of the blockchain status they care about and ignore other data unrelated to them, the entire blockchain can also operate. For new users, it is necessary to keep at least one piece of the status for determining the sub-status they care about. But the data can be stored in any number of nodes in pieces, so that the load on each node can be arbitrarily small. The Merkle tree can be used to prove that a sub-status is part of a status. Some nodes in the network are dedicated to saving trees and rewards can be gained after providing inserts and queries for these trees.

In the DSDIN network, there are many light clients that interact with users, and these clients do not need to download the entire chain. First, the user can assign his client a Hash value in the branch fork history. With this Hash value, the client knows that only the fork containing the Hash block needs to be downloaded, and only the head of the block needs to be downloaded. The head is much smaller than the entire block, so only very few transactions need to be processed. The head of a block contains:

- The Hash value of the previous block;
- Root Hashes of all status trees;

In the DSDIN network, most of the transactions are done in the wormholes, as they are independent and run in parallel with high throughput. Most of these transactions cannot be executed on the blockchain at all, or even cannot be recorded on the blockchain. In addition, the wormholes do not write any shared status to the blockchain, so all transactions recorded on the blockchain can be processed in parallel. Assume that the current mainstream common computer is equipped with a 4-core processor, then, through parallel processing, the throughput of transaction processing is magnified 4 times.

In the DSDIN network, we use the Erlang virtual machine to realize each of the wormhole, and use the Erlang process to realize each status channel. Since the Erlang process is the running mode of Actor-model, they are executed independently and in parallel, and the processes interact with each other through Message-Passing [22]. Based on the distributed nature of Erlang, each status channel can save the status it requires and maintain the branch of the transaction tree it cares about, so that the blockchain sharding can be well implemented.

*N. Virtual Machine and Distributed Computing*

DSDIN virtual machine VM is a stack-based bytecode language executor based on Erlang development. The DSDIN VM supports functions rather than goto statements, which are more advanced, more semantic, and easier to understand compared with scripting languages of Bitcoin and Ethereum. The DSDIN VM can also execute Erlang compiled bytecodes and support more advanced programming languages in subsequent versions.

Erlang is a structured and dynamic functional programming language with built-in parallel computing support. It is originally designed for communication applications by Ericsson, such as control of switches or conversion of protocols. Therefore, it is ideal for building distributed systems with real-time soft parallel computing. Applications written in Erlang typically consist of thousands of lightweight processes during operation and communicate with each other via messaging. Inter-process context switching is only one or two links for Erlang, much more efficient than thread switching of C program [21][22].

Because of the imperative and functional property, writing distributed applications with Erlang/Elixir is much more efficient [28], and the distributed mechanism is transparent: the programs do not know the distributed operation. The Erlang runtime environment is a virtual machine, so that once the code is compiled, it can be run anywhere and its runtime system even allows code to be updated without interruption, all which favor the flexible industrial applications and the development of Holonic systems in manufacturing control context [29].

### III. APPLICATIONS

*A. Wormhole Application*

In the DSDIN, wormholes are environments that provide transactions and micro-service applications with high throughput and high concurrency. The basic applications provided by wormholes include:

1. Service API
   The current mainstream API on the Internet is a HTTP-based RESTful API that provides a paradigm for interaction and development platforms between complex systems. Many APIs either obtain access rights through username-password pairs or pass the unique access Token. For some payment channel APIs, a method to charge for access by API call is provided, which can effectively resist DDoS and easily develop high quality APIs. Users are encouraged by paying for access to develop more commercially valuable APIs and realize open access to data. In the DSDIN, users can use the API and algorithm building tools provided in the wallet applications, based on high-level languages such as Javascript, R, Python or Erlang, to realize a high-quality API. The user sets the number of DSDs that the API needs to pay each time it is called through the smart contract. Since the services implemented in actor-model and run in parallel, high data throughput and service capacity could be achieved.

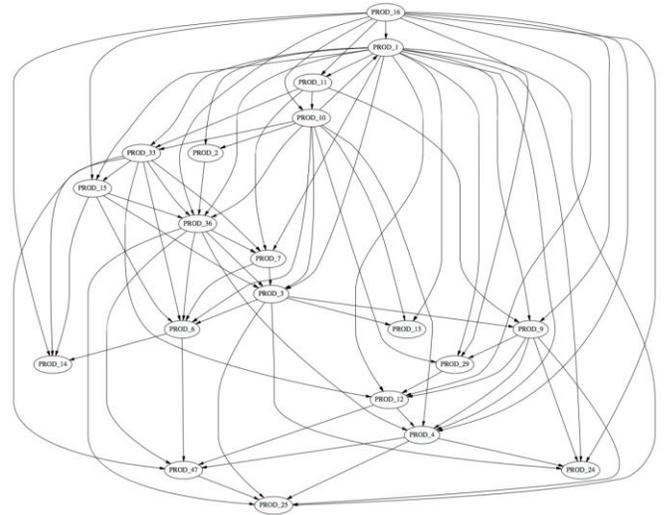

Fig. 6. Dynamic Prediction of Multi-factor Correlation and Causality based on Bayesian Network.

2. Insured crowdfunding
   In the DSDIN, insured crowdfunding can be achieved through dominant assurance contracts, which helps users to achieve crowdfunding of funds through smart contracts, such as new product design, building plants and so on. Unlike traditional centralized platform-based crowdfunding (such as JD Crowdfunding), we use the oracle to ensure that only after the crowd-funders provide the corresponding goods or services that can them receive the corresponding crowdfunding funds; if the crowdfunding fails, participants will get back their DSD along with interest.

3. Cross-chain atomic exchange
   Cross-chain atomic exchange allows non-trusted parties to directly implement the exchange of DSD and Bitcoin via a hash lock, which locks the transaction to the same value on both chains simultaneously.

4. Smart contract
   The smart contract defines that the payment is made only when an event occurs and no payment is made when the event does not occur, as determined by the oracle. Smart contracts can be used to implement insurance. For example, crops that could have been harvested turn out to be worthless due to snowstorms, but if the oracle predicts the snow, the farmers will receive the money and the investment thus can be protected. Smart contracts can also be used to motivate the opening of sensitive information, for example, using a smart contract to place a bet that "leakage will occur on AB-type lithium-ion power batteries produced by certain battery company before January 20, 2018," then anyone who first publishes the bet and releases the information will obtain an incentive from the DSD.

5. Prediction market

    The prediction market allows users to bet on events that will occur in the future. The likelihood of future events can be predicted based on the price of the chips. This is the most accurate way to measure the future at a given price. Once the event occurs, then the oracle is used to liquidate the market [30][31]. We can also use the prediction market to estimate how many products can actually be completed, so that fraud and unfounded promises can be easily detected.

### B. Intelligent Service Algorithm (ISA)

In the DSDIN, each user (enterprise or individual user) can leverage the smart contract and algorithm building tool provided by the ISA generator running on the distributed nodes, without coding or using a small amount of coding (e.g. using a high-level programming language: Javascript, R, Python, XML, etc.), and quickly build, debug, and run algorithmic models through visual module components. In the DSDIN, each algorithm model is called Intelligent Service Algorithm (ISA). All ISAs are compliant with the decentralized consensus mechanism and workload proof mechanism of the DSDIN network (The DSDIN network uses workload verification mechanism integrating PoW and PoS). ISA obtains system resources (including data, micro-services, device control rights, energy use rights, artificial intelligence algorithms, other ISAs, DSDs, etc.) through workload proof. ISA processes and analyzes real-time data from the Internet of Things, controls devices and runs applications, providing dynamic and personalized intelligent services to customers. For example, the user submits a product model to a remote 3D printer, and the printer automatically prints a fixed number of products that meet the user's needs according to the model parameters and constraints in the ISA, and delivers the product to the user. In this interaction, the user first edits an ISA using the ISA generator, and encapsulates the product 3D model, material constraints, production time constraints, and yield and device control logic. Then the user initiates a transaction based on the state channel and the current wormhole executes the ISA together with another wormhole running the 3D printer control service through the Message-Passing mechanism to implement product printing. When the printing is completed, the status channel is closed, and the economic result of the transaction is recorded on the blockchain.

### C. ISA Generator

The ISA generator is the primary application of the DSDIN wormhole, and each ISA generator can run as a separate wormhole. As shown in Fig. 12, the ISA generator is based on Erlang distributed computing technology, and serves as the basic platform for intelligent manufacturing and industrial big data applications. The underlying platform is mainly embodied in the IaaS and PaaS layers, specifically, to build a unified hardware and computing resource infrastructure environment through IaaS, and to establish unified data acquisition, aggregation, stream processing, storage, big data analysis, artificial intelligence algorithm framework and operating environment through PaaS.

### D. ISA Market

Users in the DSDIN, whether corporate or personal, are able to analyze and predict the customer needs and industrial field environment changes in a more dynamic and systematic way by leveraging the modeling capabilities, algorithm components and powerful data acquisition, aggregation, transmission and distributed computing capabilities provided by the ISA generator. The automatic optimization and adjustment of services based on analysis results, even automatically adaption to the environment, and independent decision-making can be made, thus bringing a highly personalized experience to customers.

People (including enterprise users or individual users) can get Token motivation from the DSDIN reward pool by building models, designing algorithms, contributing domain knowledge, or providing computing power through edge computation on the DSDIN, or by providing data through decentralized storage on the DSDIN. The algorithm market platform freely opens some basic algorithm models and service components, as well as limited platform computing power to users (we refer to them collectively as basic resources, abbreviated as BR). Among them, the basic algorithm models include: basic statistics (such as the R packages) and ML libraries etc.; basic service components include: Data access and transmission (MQTT, TCP, HTTP, Websocket, SQL, File, Excel, etc.); platform computing power is determined by several factors, including: Data source access concurrency, data access throughput, database type, data storage, computing speed, number of distributed computing nodes, etc. Other advanced algorithm models, advanced service components, algorithm models or service components provided by third-party users, and platform powers above a certain threshold (collectively referred to as advanced resources, abbreviated as AR), shall be purchased through the DSD of the platform.

The ISA market provides the DSDIN network with a rich set of data, data models, artificial intelligence algorithms, service components and intelligent services for different industry services and applications. The algorithm store is open to third-party developers, and developers can sell some kind of platform resources (including those mentioned above: Data, data models, artificial intelligence algorithms, service composition and intelligent services, etc.), or obtain Token revenue by realizing resources allocation.

### E. ISA Zone

The wormhole application of the DSDIN provides an innovative Algorithm Zone (AZ), each of which is an independent industrial intelligence network platform for a single industry, a single enterprise (group), a single field, or even an individual. Any platform user (enterprise or individual) can pay DSD to create an AZ. Each AZ has a unique access ID (AZID), which facilitates access and identifying of AZ members. AZ ownership will be recorded through the blockchain account to ensure the authenticity and tamper resistance. AZ can be freely traded on the market. The user who creates the AZ is the AZ owner and can specify one or more domain administrators for its AZ. The AZ owner can decide whether to pay the Token and how many Tokens need to be paid in joining AZ. Users on the DSDIN can apply to join one or more zones, but for each application a smart contract provided by the corresponding AZ shall be signed and the application needs to be recorded into the blockchain ledger. When the application conditions are met (such as paying the corresponding Token or contributing an algorithm or knowledge), the smart contract will be automatically fulfilled, and the user becomes the member of AZ.

Members within AZ can use the basic resources in the zone to generate (access) data, process and analyze data, create algorithms, build intelligent services, and publish status, follow each other, share their own data, analysis results, created algorithms or intelligent services, and each member's status will be instantly displayed on the AZ internal timeline.

In order to promote data sharing, knowledge sharing, algorithm and intelligent service sharing, the platform provides cross-zone interaction mechanism, and users can apply to join multiple AZs, and can recommend specific users to specific zones, or specific zones to specific users. If a user "refers" another user to an AZ domain, the "referral" user can browse the status on the timeline of the zone, but cannot post status, nor create algorithms, data, services, etc. unless the user applies for joining and officially joins the zone.

### F. Wormhole based Equipment Service Management

The devices involved in the DSDIN for ISA trading mainly include production lines, machine tools, energy-consuming equipment, industrial personal computers and sensors, 3D printers, etc., and these devices also generate IoT data in real time, which is stored in the decentralized storage system of wormhole management. Based on the time series nature of IoT data, we can abstract most of the prediction or detection problems into a Semi-Markov Decision Process[32][33]. For a group of devices that contain N identical devices, assuming that there are M states for each device, then the total device configuration is MN. Here we are not simply predicting for prediction, but directly optimize goals in the production process by reinforcement learning algorithms, e.g. maximizing the service capabilities of the above device group, then the decision problem will be defined as: At time $t$, the system gets an immediate return $r_t$, defined as the number of service requests to the device group at that time; the system return is:

$$R(t) = \int_0^\infty \exp(-\beta\tau) \ r_{t+1} d\tau. \quad (1)$$

$\beta > 0$ is the discount factor of the reward here. To maximize the expected value of return is the primary goal of real-time data processing for this project, which is equivalent to minimizing the number of times a device group cannot respond to a service request within the limited lifetime of the device.

For the above optimization decision, the wormhole application is mainly based on the Temporal-Difference algorithm (TD Learning), which is also optimized with plenty of parallel processes. The main algorithms are:

1) On-Policy TD Control

The first is to learn the action-Value function, to estimate the state value function for the current behavior policy π, all states s and action a $Q^\pi(s,a)$, and to divide the learning process into multiple episodes, thus forming an interleaved sequence of state and state-action pair. The iterative process of the Q function is:

$$Q(s_t, a_t) \leftarrow Q(s_t, a_t) + \alpha[(r_t + \gamma Q(s_{t+1}, a_{t+1}) - Q(s_t, a_t)] \quad (2)$$

2) Off-Policy TD Control

The iterative process of the Q function is:

$$Q(s_t, a_t) \leftarrow Q(s_t, a_t) + \alpha[(r_t + \gamma max_a Q(s_{t+1}, a_{t+1}) - Q(s_t, a_t)] \quad (3)$$

Meanwhile, the Actor-Critic method will be adopted for some processes required to be with precise and demanding feedback on production control. The learning process is always on-policy, and the TD error is strictly controlled.

On the other hand, the related device groups or related processes will undergo correlation and probability distribution prediction based on Prediction Market with Bayesian Networks. As the integration complexity of edge conditional probabilities will increase exponentially with the number of variables, the platform will allow each Erlang process to act as a watching node fully based on Erlang's Actor-Model feature, thereby implementing Belief Propagation (BP) via Message-Passing. Besides, because the Erlang process is inherently distributed and it can run on the same Erlang VM or on different VMs, or even on different computer nodes, so the BP algorithm is extremely scalable.

### G. Smart Contract

Smart contracts in the DSDIN exist in wormholes and are implemented through status channels, which enable the perfect implementation of micro-services featuring high data throughput in the Industrial Internet of Things (IIoT). Besides, the DSDIN smart contract is Turing-complete and can be implemented using a functional high-level programming language, including Javascript, Erlang, R, Python. In the DSDIN ecosystem, many constraints and transactions will be executed through smart contracts. The following sections describe some of the contract implementations associated with the ISA platform.

1. AZ information
   On the ISA platform, anyone or organization can create their own AZ by paying a certain amount of DSDs. To indicate the uniqueness and ownership of the AZ, the zone information and its attribution will be written to the DSDIN smart contract. The contract type includes creating and purchasing a zone, and all platform resources, user information, and status of the zone.
2. Algorithm or service
   On the ISA platform, any user can create a series of data, algorithms or services based on the BRs provided by the platform or other ARs, collectively referred to as resources. To indicate the uniqueness and ownership of these algorithms or services, the resource information and its attribution will be written to the DSDIN smart contract. Contract types include creating, publishing and purchasing resources, and all information and status of the resources are included in the contract.
3. Ecological rules
   Incentive rules of reward pool; The proportion of revenue generated by the ISA platform and third-party developers through resource sharing; The number of DSDs required for creating an AZ; And other ecological rules that need to be added as the ISA platform evolves.

## IV. MOTIVATION MECHANISM

DSDIN has set up a special eco-reward fund of ISA platform, which uses DSDIN's native Token (DSD) as the sole reward method. The platform employs a unique POV (Proof of Value) + POC (Proof of Contribution) algorithm to establish a double reward mechanism for rewarding participants who contribute to the building of ISA community, as depicted in Fig. 7. The reward pool system performs the incentive allocation calculation every 72 hours, and automatically distributes the reward Token to the participant's personal wallet based on the calculation result.

### A. Rewarding Pool

The reward pool is a Token pool dedicated to rewarding eco-construction contributors for DSDIN industrial intelligence network. The eco-incentive pool automatically replenishes the reward pool every 72 hours according to the incentive value function Q(t). The iteration of each Q(t) is as follows:

$$Q(t) \leftarrow Q(t) + \alpha[\gamma(t) + \mu Q(t+1) - Q(t)] \quad (4)$$

Where Q(t) is the value function of current incentive pool, Q(t+1) the next 72-hour value function, α the step size coefficient, $\gamma(t)$ the currently rewarded Token number, and μ the correction factor.

### B. AZ Reward Distribution – Proof of Value

The reward pool uses the POV (Proof of Value) algorithm to perform the first layer of reward distribution, that is, the DSD in the reward pool is allocated to each AZ according to the value of the AZ.

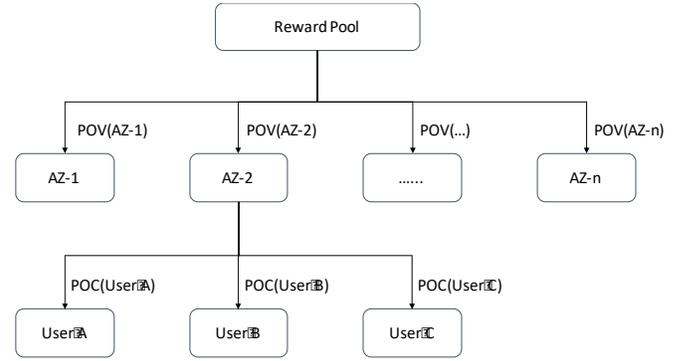

Fig. 7. Double-layer Incentive Mechanism for ISA Ecosystem.

Calculation methods of POV as follows: The reward pool calculates, every 72 hours, valid data, the number of algorithms, the number of effective services, the number of service components, the total number of times effective algorithms are used, the total number of times the service components are used, the total duration of effective services, user activity, the number of releases of effective content, the average online duration of users, and the number of new members for each AZ, the value V(n) of each AZ is:

$$V(n) = \sum_{i=1}^{k} \alpha_i F_i$$

Where $k$ is the total number of factors, and $F_i$ is the normalized value of each factor within the AZ, calculated according to the corresponding value function. $\alpha_i$ the weight of different factors in the zone value.

The number of new DSD Tokens added to the reward pool every 72 hours is γ; all AZs on the platform are allocated and the reward $\gamma_n$ for each AZ as a whole is:

$$\gamma_n = \gamma * \frac{V_n}{\sum_{i=1}^{m} V_i}$$

Where $m$ is the number of AZs within the ecosystem of the entire ISA platform as of the time of allocation.

### C. User Reward Distribution – Proof of Contribution

The member users in each AZ will be assigned the overall reward $\gamma_n$ from the reward pool according to the contribution of the work based on the POC (Proof of Contribution) algorithm.

DSDIN calculates a user's POC $W_n$ through the work and published content of such user in the AZ, and the usage of the published content. The factors determining a user's contribution proof include: The number of valid ISAs created by the user, the number of service components, the number of valid services, the usage of the work content, and the content published in the AZ and activity level.

$$W_n = \varepsilon * \sum_{i=1}^{k} \alpha_i S_i + \theta * \sum_{i=1}^{k} \beta_i \sum_{j=1}^{m} C_j$$

In which, $S_i$ is the normalized value of the user's POC single work content factor calculated based on the contribution value function, $\alpha_i$ is the weight of each work content factor, and ε is the value weight of the user's active contribution content; $C_j$ is the normalized value of a user's contribution content that is used by others at a certain time, and the value of usage is determined by the duration of use and the value of the service generated by the use on the platform; $\beta_i$ is the usage weight of each contribution content, and θ is the value weight of user's work content that is used; *k* is the number of work content contributed by the user, and *m* is the number of times that a work content is used.

Every 72 hours, the number of DSDs received by the AZ that the ISA platform rewards through the POV mechanism is γ, and these DSDs will be rewarded by the POC mechanism to users who contribute to the AZ within 72 hours. The reward that user *n* receives for the AZ within 72 hours is:

$$\gamma_n = \gamma * \frac{W_n}{\sum_{i=1}^{m} W_i}$$

Where, *m* is the total number of the AZ members.

## V. Conclusion

This paper has presented how to architect a globally distributed industrial network for Industrial Internet of Things and demonstrated how value is transferred across this network for achieving intelligent services as the fundament of Industrial 4.0. Particularly the applications introduced in section III are easy to be built and can efficiently and effectively work on DSDIN. The motivation mechanism introduced in section IV will make a healthy and promising value creation ecosystem, which encourages participants to contribute and form a positive feedback network, effectively transferring values.